\documentclass[prl, aps, twocolumn, floatfix, showpacs]{revtex4}
\usepackage{graphicx, amsmath, amssymb, times}

\begin{document}
\title{Fidelity susceptibility and topological phase transition of a two
dimensional spin-orbit coupled Fermi superfluid}
\author{Xuebing Luo$^{1}$}
\author{Kezhao Zhou$^{1,2}$}
\email{kzzhou@imr.ac.cn}
\author{Wuming Liu$^{3}$}
\author{Zhaoxin Liang$^{1}$}
\author{Zhidong Zhang$^{1}$}
\affiliation{$^1$Shenyang National Laboratory for Materials Science, Institute of Metal
Research, Chinese Academy of
Sciences, 72 Wenhua Road, Shenyang 110016, China\\
$^2$Department of Physics, University of Illinois, 1110 W. Green Street,
Urbana, IL 61801-3080\\
$^3$Institute of Physics, Chinese Academy of Sciences, Beijing, 100080,
China }

\date{\today}

\begin{abstract}
We investigate the fidelity susceptibility (FS) of a two-dimensional
spin-orbit coupled (SOC) Fermi superfluid and the topological phase
transition driven by a Zeeman field in the perspective of its ground-state
wave-function. Without Zeeman coupling, FS shows new
features characterizing the BCS-BEC crossover induced by SOC. In the
presence of a Zeeman field, the topological phase transition is explored using
both FS and the topological invariant. In particular, we obtain the
analytical result of the topological invariant which explicitly demonstrates
that the topological phase transition corresponds to a sudden change of the
ground state wave-function. Consequently, FS diverges at the phase
transition point with its critical behavior being: $\chi \propto ln|h-h_{c}|$
. Based on this observation, we conclude that the topological phase
transition can be detected by measuring the momentum distribution in cold
atoms experiment.
\end{abstract}

\maketitle

Spin-orbit coupling (SOC) is a key ingredient in realizing nontrivial
topological phases which is of great interest in current physics community
\cite{nayak, qi, ber}. Particularly, semiconductors with proximity-induced
superconductivity and external effective Zeeman coupling as well as
non-centrosymmetric superconductors provide real material examples with a
non-Abelian quantum order \cite{sarma}. Another promising platform of
achieving this scenario is ultra-cold Fermi systems where SOC, Zeeman
coupling and superconductivity can be readily realized in current
experimental set-up \cite{spielman, zhang, cheuk} and construction of model systems on the Hamiltonian
level is now available \cite{dalibard, goldman}. Furthermore, two dimensional (2D)
geometry can be created by generating a strong trapping potential along one
direction \cite{mart, fro, feld}.

Theoretical investigations have shown that SOC has nontrivial effects on
various properties of the Fermi superfluid systems. In the absence of Zeeman
coupling, SOC can produce a novel bound-state called Rashbons \cite{shenoy, huhui}
and therefore induce a crossover from weakly correlated BCS to strongly
interacting BEC regime (BCS-BEC) even for very weak particle-particle
interaction \cite{yu, han, he, sala}. Of particular interest is the topological
phase transition driven by a Zeeman field \cite{iskin, gong, wu, duan} which can
be classified by a topological invariant constructed for this particular
scenario \cite{sarma} and also in He$^{3}$-A \cite{volovic} where its
theoretical structure and relation with other invariants have been discussed
comprehensively. However, much of the information encoded in the topological invariant
has been buried in the numerical procedure and therefore analytical result
is of great interest and value for the full understanding of these novel states. Both
of these two aspects are fundamentally related to the ground-state wave-function of the many-body
systems considered. As a direct measure of the change of the ground-state wave-function, fidelity susceptibility (FS)
has been used extensively to study the quantum phase transition problems \cite{shijian} which is the main motivation of this paper.

In this Letter, we consider a typical 2D s-wave Fermi superfluid in the presence of both SOC
and Zeeman coupling. In the absence of Zeeman coupling, with the
self-consistent solution of the gap and number equations, we investigate the
behaviors of fidelity susceptibility (FS) as functions of both interaction and SOC. Numerical results
show that FS exhibits a local peak structure characterizing the BCS-BEC
crossover induced by SOC which is quite different from other thermodynamic
quantities. In the presence of Zeeman coupling, we focus on situation with a
strong enough SOC such that only topological phase transition occurs. In
particular, we obtain the analytical result of the topological invariant
which provides new insights into the topological nature of the ground-state:
(I) singular behavior of the Berry curvature is determined by SOC through
the vortex-like solutions of the Bogoliubov-de Gennes (BdG) Hamiltonian
instead of the Dirac point in the excitation spectrum acting as a magnetic
monopole in momentum space; (II) the topological phase transition
corresponds to a sudden change of the ground-state wave-function at zero
momentum which is also captured by the divergence of FS.
Finally, its critical behavior is obtained by analyzing the gapless
excitation spectrum at the phase transition point.

\textit{Formalism.}---The system under consideration can be described by
the Hamiltonian: $H=H_{0}-g\int d^{2}\mathbf{r}\varphi _{\uparrow }^{\dagger
}\left( \mathbf{r}\right) \varphi _{\downarrow }^{\dagger }\left( \mathbf{r}
\right) \varphi _{\downarrow }\left( \mathbf{r}\right) \varphi _{\uparrow
}\left( \mathbf{r}\right) $ with $g>0$ being the contact interaction
parameter and $\varphi _{\sigma \left( =\uparrow ,\downarrow \right) }\left(
\mathbf{r}\right) $ and $\varphi _{\sigma }^{\dagger }\left( \mathbf{r}
\right) $ are the annihilation and creation field operators, respectively.
The non-interacting part $H_{0}$ can be written as $H_{0}=\int d^{2}\mathbf{
r }\psi ^{\dagger }\left( \mathbf{r}\right) \left[ \varepsilon _{\mathbf{
\hat{p }}}-h\sigma _{z}+\lambda \left( \mathbf{\ \sigma }\times \mathbf{\hat{
p}} \right) \cdot \mathbf{z}\right] \psi \left( \mathbf{r}\right) $ where $
\psi \left( \mathbf{r}\right) =\left[ \varphi _{\uparrow }\left( \mathbf{r}
\right) ,\varphi _{\downarrow }\left( \mathbf{r}\right) \right] ^{T}$ and
kinetic energy $\varepsilon _{\mathbf{\hat{p}}}=\mathbf{\hat{p}}^{2}/2m-\mu $
with $m$, $\mu $\ and $h$ being the mass of the Fermi atoms, the chemical
potential and the effective Zeeman field, respectively. For simplicity we
set $\hbar =1$ throughout this Letter. The third term is the Rashba SOC \cite{rashba} with
$\lambda >0$ denoting the strength of SOC and $\mathbf{\sigma }$ being the
Pauli matrices. Within mean-field theory, the interacting part can be
approximated by $H_{BCS}=-\int d^{2}\mathbf{r}\left( \Delta \left( \mathbf{r}
\right) \varphi _{\uparrow }^{\dagger }\left( \mathbf{r}\right) \varphi
_{\downarrow }^{\dagger }\left( \mathbf{r}\right) +h.c.\right) +\int d^{2}
\mathbf{r}\left\vert \Delta \left( \mathbf{r}\right) \right\vert ^{2}/g$
with $\Delta \left( \mathbf{r}\right) $ being the pairing field.

This model Hamiltonian $H=H_{0}+H_{BCS}$ is also relevant to semiconductors
where superconductivity is induced by proximity effect and
non-centrosymmetric superconductors in the sense that p-wave pairing
component does not affect much the topological properties of the system.
As shown in Ref. \cite{sarma}, the Majorana zero-energy state for the vortex
solution of BdG equation contains the essential details of the nontrivial
topological nature of the ground-state. For our purpose, we only consider
translational invariant solutions where the paring field becomes a constant $
\Delta \left( \mathbf{r}\right) =\Delta $. Consequently, in momentum space,
the total Hamiltonian reduces to $H=\sum_{\mathbf{p}>0}\Phi _{ \mathbf{p}
}^{\dagger }H_{BdG}\left( \mathbf{p}\right) \Phi _{\mathbf{p} }+\sum_{
\mathbf{p}}\varepsilon _{\mathbf{p}}+V\Delta ^{2}/g$ where $V$ denotes the
size of the system, $\Phi _{ \mathbf{p}}=\left[ a_{\mathbf{p},\uparrow },a_{
\mathbf{p},\downarrow },a_{-\mathbf{p},\uparrow}^{\dagger },a_{-\mathbf{p}
,\downarrow}^{\dagger }\right]^{T} $ and the BdG Hamiltonian $H_{BdG}\left(
\mathbf{p}\right) $ is
\begin{equation}
H_{BdG}\left( \mathbf{p}\right) =\left[
\begin{array}{cccc}
\varepsilon _{\mathbf{p}}-h & \Gamma _{\mathbf{p}} & 0 & -\Delta \\
\Gamma _{\mathbf{p}}^{\ast } & \varepsilon _{\mathbf{p}}+h & \Delta & 0 \\
0 & \Delta & -\varepsilon _{\mathbf{p}}+h & \Gamma _{\mathbf{p}}^{\ast } \\
-\Delta & 0 & \Gamma _{\mathbf{p}} & -\varepsilon _{\mathbf{p}}-h
\end{array}
\right]  \label{ham}
\end{equation}
with $\Gamma_{\mathbf{p}}=\lambda \left(p_{y}+ip_{x}\right)$. Using the
standard diagonalization procedure, we obtain the ground-state free energy $
E_{g}=\sum_{\mathbf{p},s\mathbf{=\pm}}\left(\varepsilon_{\mathbf{p}}-E_{
\mathbf{p},s}\right)/2+V\Delta^{2}/g$ where the excitation spectrum $E_{
\mathbf{p},s}=\sqrt{\mathcal{E}_{ \mathbf{p},s}^{2}+\Delta _{\mathbf{p}
,2}^{2}}$ with $\mathcal{E}_{\mathbf{p}, s}=E_{\mathbf{p}}-s\sqrt{
h^{2}+\left\vert\Gamma_{\mathbf{p}}\right\vert ^{2}}$, $E_{\mathbf{p}}=\sqrt{
\varepsilon_{\mathbf{p}}^{2}+\Delta_{\mathbf{\ p},1}^{2}}$, $\Delta_{\mathbf{
\ p},1}=\Delta\left\vert\cos\theta_{\mathbf{p}}\right\vert$, $\Delta_{
\mathbf{p },2}=\Delta\sin\theta_{\mathbf{p}}$ and $\theta_{ \mathbf{p}
}=\pi-\arctan\left(\left\vert\Gamma_{\mathbf{p}}\right\vert /h\right)$. For
fixed $h$, $\lambda$ and $g$, $\Delta$ and $\mu$ are given by the
self-consistent solutions of the gap and number equations
\begin{eqnarray}
\frac{1}{g} &=&\frac{1}{V}\sum_{\mathbf{p},s}\frac{1+s\cos\theta_{\mathbf{p}
} \frac{h}{E_{\mathbf{p}}}}{4E_{\mathbf{p},s}}, \label{gap} \\
N &=&\frac{1}{2}\sum_{\mathbf{p},s}\left( 1-\frac{\mathcal{E}_{\mathbf{p},s}
}{E_{\mathbf{p},s}}\frac{\varepsilon _{\mathbf{p}}}{E_{\mathbf{p}}}\right).
\label{number}
\end{eqnarray}

Note that in the presence of Zeeman coupling, Eq. ({\ref{gap}})\ generally
supports more than one solution, while the physical one corresponds to the
global minimum point of $E_{g}$. As usual, divergence of the integral over
momenta in Eq. ({\ref{gap}}) is removed by replacing contact interaction
parameter $g$ by binding energy $E_{b}$ through $V/g=\sum_{\mathbf{p}
}1/\left( 2\epsilon _{\mathbf{p}}+E_{b}\right) $.

Furthermore, the ground-state wave-function can be directly given by the
unitary transformation that diagonalizes Eq. ({\ref{ham}}). However, we find
that the final result becomes more physically transparent by using
Bogoliubov transformation step by step. First, in helicity basis: $c_{
\mathbf{p},s}=\sin\left(\theta_{\mathbf{p}}/2\right)a_{\mathbf{p},s}-s\cos
\left(\theta_{\mathbf{p}}/2\right)e^{is\varphi_{\mathbf{p}}}a_{\mathbf{p}
,-s}$ with $\varphi_{\mathbf{p}}=\arctan\left(p_{x}/p_{y}\right)$, the
total Hamiltonian becomes $H=V\Delta ^{2}/g+\sum_{\mathbf{p},s}\xi _{\mathbf
{p},s}c_{\mathbf{p},s}^{\dagger}c_{\mathbf{p},s}-1/2\sum_{\mathbf{p},s}
\left(\Delta_{\mathbf{p},2}e^{is\varphi_{\mathbf{p}}}c_{\mathbf{p},
s}^{\dagger }c_{-\mathbf{p},s}^{\dagger}-s\Delta_{\mathbf{p},1}c_{\mathbf{
p},s}^{\dagger}c_{-\mathbf{p},-s}^{\dagger}+h.c.\right) $ with $\xi _{
\mathbf{p},s}=\varepsilon _{\mathbf{p}}+s\left\vert \Gamma _{\mathbf{p}
}\right\vert$ from which we see that pairing happens between both the same
and different helicity basses. Second, using Bogoliubov transformation: $
\beta_{\mathbf{p},s}=u_{\mathbf{p}}c_{\mathbf{p},s}-v_{\mathbf{p}}c_{-
\mathbf{p},-s}^{\dagger}$ with $u_{\mathbf{p}}=\sqrt{\left(1+\varepsilon
_{\mathbf{p}}/E_{\mathbf{p}}\right)/2}$ and $u_{\mathbf{p}}^{2}+v_{\mathbf{
p}}^{2}=1$, the Hamiltonian reduces to the standard pairing form as: $
H=V\Delta ^{2}/g+\sum_{\mathbf{p} }\left( \varepsilon_{\mathbf{p}} -E_{
\mathbf{p}}\right) +\sum_{\mathbf{p},s}\mathcal{E}_{ \mathbf{p},s}\beta _{
\mathbf{p},s}^{\dagger }\beta _{\mathbf{p},s}-1/2 \sum_{\mathbf{p},s}\Delta
_{\mathbf{p},2}\left[ e^{is\varphi _{\mathbf{p} }}\beta _{\mathbf{p}
,s}^{\dagger }\beta _{-\mathbf{p},s}^{\dagger }+h.c. \right] $. Finally, the
ground-state wave-function can be easily obtained as
\begin{equation}
\left\vert G\right\rangle =\prod\limits_{\mathbf{p}>0,s}\left( u_{ \mathbf{p}
,s}+e^{is\varphi _{\mathbf{p}}}v_{\mathbf{p},s}\beta _{\mathbf{p},
s}^{\dagger }\beta _{-\mathbf{p},s}^{\dagger }\right) \left\vert
g\right\rangle  \label{wavefunction}
\end{equation}
where $\left\vert g\right\rangle =\prod\limits_{\mathbf{p}}\left( u_{\mathbf{
p}}+v_{\mathbf{p}}c_{\mathbf{p},+}^{\dagger }c_{-\mathbf{p},- }^{\dagger
}\right) \left\vert 0\right\rangle $, $u_{\mathbf{p},s}$ and $v_{\mathbf{p}
,s}$ are given as
\begin{equation}
\left[
\begin{array}{c}
u_{\mathbf{p},s} \\
v_{\mathbf{p},s}
\end{array}
\right] =\sqrt{\frac{1}{2}\left( 1\pm \frac{\mathcal{E}_{\mathbf{p},s}}{E_{
\mathbf{p},s}}\right) } . \label{uvs}
\end{equation}
Here, $\left\vert g\right\rangle $, constructed to be the vacuum state of $
\beta _{ \mathbf{p},s}$, can be considered as singlet pairing of different
helicity states ($c_{ \mathbf{p},s}$). The ground-state wave-function $
\left\vert G\right\rangle $ describes triplet pairing of the quasi-particles
denoted by $\beta _{ \mathbf{p},s}$. In the absence of Zeeman coupling, $u_{
\mathbf{p}}=1$, $v_{ \mathbf{p}}=0$ and $\left\vert g\right\rangle
=\left\vert 0\right\rangle $, $\beta _{\mathbf{p},s}$ is the helicity basis
and $\left\vert G\right\rangle $ represents state with pairing only happening
in the same helicity basis \cite{yu}.

Finally, as a direct measure of the change of ground-state wave-function, FS
is defined as the following form:
\begin{equation}
\chi \left( \alpha \right) =\left\langle G\right\vert \overleftarrow{
\partial }_{\alpha }\partial _{\alpha }\left\vert G\right\rangle
-\left\langle G\right\vert \overleftarrow{\partial }_{\alpha }\left\vert
G\right\rangle \left\langle G\right\vert \partial _{\alpha }\left\vert
G\right\rangle  \label{fidelity}
\end{equation}
where $\left\vert G\right\rangle $ denotes the ground-state wave-function of
the many body systems and $\alpha $ is the control parameter. Recent
investigations show that it provides an effective way of determining the
phase transition boundary \cite{shijian}. The critical behaviors of FS near
quantum phase transition are of great interest especially for topological
phase transitions. On the other hand, it has also been used to study the
crossover induced by interaction in the absence of SOC with its width being
associated with the crossover region \cite{khan}.

\textit{Balanced case.}---For balanced case $h=0$, self-consistent solution
of the gap and number equations gives $\mu /E_{F}$ and $\Delta /E_{F}$ as
functions of $\tilde{\lambda} =m\lambda /k_{F}$ and $\eta =E_{B}/E_{F}$ with $E_{F}=k_{F}^{2}/2m$ being the Fermi
energy and the $k_{F}$ being defined through $k_{F}^{2}=2\pi n$. Based on these results, we investigate effect
of SOC on the behaviors of the FS in the BCS-BEC crossover problem. By
direct substitution of Eq. ({\ref{wavefunction}}) into Eq. ({\ref{fidelity}}
), we obtain
\begin{eqnarray}
\chi \left( \lambda \right) &=& \sum_{\mathbf{p},s}\frac{1}{8E_{\mathbf{p},
s}^{4}}\left[ \Delta \left( sp_{\bot }-\frac{\partial \mu }{\partial \lambda
} \right) -\xi _{\mathbf{p},s}\frac{\partial \Delta }{\partial \lambda }
\right] ^{2}, \label{fsblam}\\
\chi \left( g\right) &=& \sum_{\mathbf{p},s}\frac{1}{8E_{\mathbf{p},s}^{4}}
\left[ \Delta \frac{\partial \mu }{\partial g}+\xi _{\mathbf{p},s}\frac{
\partial \Delta }{\partial g}\right] ^{2}  \label{fsbg}
\end{eqnarray}
with $p_{\bot }=\sqrt{p_{x}^{2}+p_{y}^{2}}$.

\begin{figure}[tbh]
\includegraphics[width=\columnwidth,height=80mm]{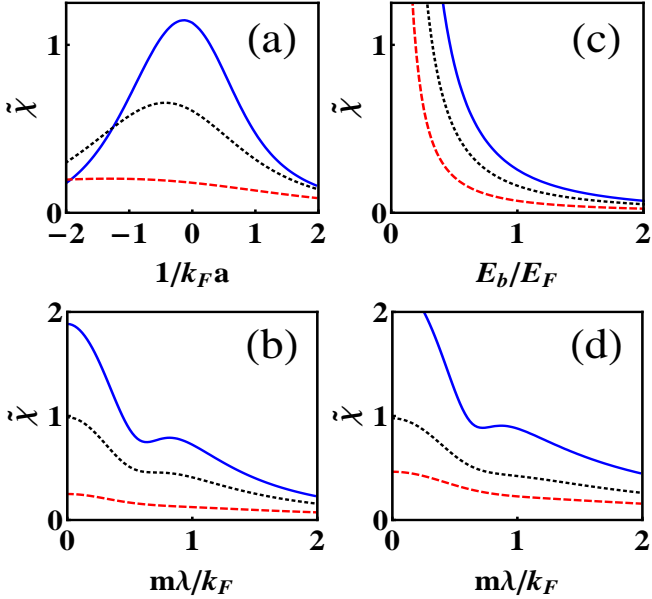}
\caption{(Color online) Fidelity susceptibility as functions of interaction
parameters and strength of SOC in three dimensions (a, b) and two dimensions
(c, d). In (a) and (c), $\widetilde{\protect\lambda}$ are given as: 0, 1 and
2 for solid blue line, black dotted and red dashed line, respectively.
Interaction parameter are set to be $\protect\eta=1/(k_{F}a)=-1.2,-1,-0.6$
in (b) and $\protect\eta=E_{b}/E_{F}=0.2,0.8,2.2$ in (d) for lines from
above. }
\label{fdlt_h0}
\end{figure}

The numerical results of Eq. ({\ref{fsblam}}) and Eq. ({\ref{fsbg}}) are
presented in Fig. (\ref{fdlt_h0}) where results in three dimensions (3D) are
also included as it provides the essential features of FS in the BCS-BEC crossover problem.
In 3D, the interaction is characterized by the scattering length
$a$ introduced by substituting $V/g=-mV/4\pi a+\sum_{\mathbf{p}}1/\left( 2\epsilon _{\mathbf{p}
}\right)$ into Eq. (\ref{gap}) and Fermi momentum is defined by $k_{F}^{3}=3\pi ^{2}n$.

For 3D case, as can be seen from Fig. \ref
{fdlt_h0} (a), there is a global peak in FS as a function of interaction
parameter $1/k_{F}a$ and it disappears for strong enough SOC. Besides, it is no longer
symmetric for finite strength of SOC. However, Fig. \ref{fdlt_h0} (b) shows
that around the critical value of $\lambda $ FS has a local maximum which
marks the crossover induced by SOC and also disappears for large enough
interaction parameters. For 2D case, as presented in Fig. \ref{fdlt_h0} (c)
and (d), this feature becomes more distinct in the sense that FS
as a function of $E_{B}$ does not show a clear crossover signature but it
does with varying SOC. Finally, it is worthwhile to note that the local peak
of FS as function of $\lambda $ is not located at the same points where
the gap parameter increases
suddenly or superfluid density has a minimum value \cite{yu,zhou}. In summary, FS provides a new different angle to
investigate the novel effects of SOC in the BCS-BEC crossover problem.

\textit{Imbalanced case}.---Here, we only focus on the topological phase
transition in 2D. For weak SOC, there will be a first order phase transition
with increasing $h$ and FS is apparently divergent due to the trivial sudden
change of $\Delta $ and therefore the wave-function. For strong enough SOC, $
\Delta $ is always nonzero even for very large $h$ and decreases
continuously with increasing $h$ \cite{duan}. However, there is a
topological phase transition across $h=h_{c}=\sqrt{\mu ^{2}+\Delta ^{2}}$.
For $h<h_{c}$, the system is in a trivial gapped superfluid state. When $
h>h_{c}$, the ground state is nontrivial and is characterized by a nonzero
topological invariant $\mathcal{N}$ which is defined as \cite{sarma,volovic}
$\mathcal{N}=1/2\pi \int_{-\infty }^{+\infty }d^{2}\mathbf{p}B\left( \mathbf{
\ p}\right) $ where the Berry curvature is given by
\begin{equation}
B\left( \mathbf{p}\right) =-i\sum_{E_{\mathbf{p}}^{\alpha }<0}\left[
\partial _{p_{x}}\mathbf{u}_{\mathbf{p},\alpha }^{\dagger }\partial _{p_{y}}
\mathbf{u}_{\mathbf{p},\alpha }-\partial _{p_{y}}\mathbf{u}_{\mathbf{p}
,\alpha }^{\dagger }\partial _{p_{x}}\mathbf{u}_{\mathbf{p},\alpha }\right]
\label{bc}
\end{equation}
with $\mathbf{u}_{\mathbf{p},\alpha =1,2,3,4}$ being the eigenvectors of Eq.
( {\ref{ham}}) corresponding to the eigenvalues $-E_{\mathbf{p},+},E_{
\mathbf{p},+},-E_{\mathbf{p},-},E_{\mathbf{p},-}$, respectively. It is clear
that the ordering and sign of the elements of the eigenvector do not
influence the Berry curvature and therefore $\mathcal{N}$. Furthermore,
only negative eigenvectors appears in $\mathbf{B}\left( \mathbf{p}\right) $.
Therefore, we can re-write the eigenvector in more transparent form as $\mathbf{
u }_{\mathbf{p},s=\pm }=\left[ e^{is\varphi _{\mathbf{p}}}F_{\mathbf{p}
,s}^{1},F_{\mathbf{p},s}^{2},F_{\mathbf{p},s}^{3},e^{is\varphi _{\mathbf{p}
}}F_{\mathbf{p},s}^{4}\right] ^{T}$ with
\begin{eqnarray*}
F_{\mathbf{p},s}^{1} &=&u_{\mathbf{p}}\sin \frac{\theta _{\mathbf{p}}}{2}v_{
\mathbf{p},s}-v_{\mathbf{p}}\cos \frac{\theta _{\mathbf{p}}}{2}u_{\mathbf{p}
,s} \\
F_{\mathbf{p},s}^{2} &=&u_{\mathbf{p}}\cos \frac{\theta _{\mathbf{p}}}{2}v_{
\mathbf{p},s}+v_{\mathbf{p}}\sin \frac{\theta _{\mathbf{p}}}{2}u_{\mathbf{p}
,s} \\
F_{\mathbf{p},s}^{3} &=&u_{\mathbf{p}}\sin \frac{\theta _{\mathbf{p}}}{2}u_{
\mathbf{p},s}+v_{\mathbf{p}}\cos \frac{\theta _{\mathbf{p}}}{2}v_{\mathbf{p}
,s} \\
F_{\mathbf{p},s}^{4} &=&u_{\mathbf{p}}\cos \frac{\theta _{\mathbf{p}}}{2}u_{
\mathbf{p},s}-v_{\mathbf{p}}\sin \frac{\theta _{\mathbf{p}}}{2}v_{\mathbf{p}
,s}.
\end{eqnarray*}

Simple algebraic manipulation gives
\begin{equation}
B\left( \mathbf{p}\right) =\partial _{p_{y}}\phi _{\mathbf{p}}\partial
_{p_{x}}F_{\mathbf{p}}-\partial _{p_{x}}\phi _{\mathbf{p}}\partial
_{p_{y}}F_{\mathbf{p}}=-\frac{\mathbf{p}}{\left\vert \mathbf{p}\right\vert
^{2}}\cdot \mathbf{\nabla }F_{\mathbf{p}}  \label{berry2D}
\end{equation}
and $F_{\mathbf{p}}=\sum_{\alpha=1,4 ,s}s\left( F_{\mathbf{p},s}^{\alpha
}\right) ^{2}$. $\mathcal{N}$ can now be easily obtained as
\begin{equation}
\mathcal{N}=F_{\mathbf{0}}=v_{\mathbf{0},+}^{2}=\theta \left( h-h_{c}\right).
\label{invariant}
\end{equation}

\begin{figure}[tbh]
\includegraphics[width=\columnwidth,height=60mm]{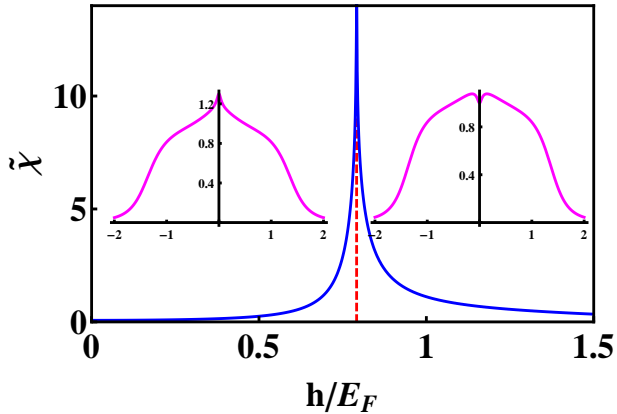}
\caption{(Color online) Fidelity susceptibility as functions of Zeeman field
with $E_{b}/E_{F}=0.5$ and $m\protect\lambda/k_{F}=1$. The vertical red
dashed line denotes the critical Zeeman field $h_{c}=0.792$. The two
inserted figures correspond to the momentum distribution of the total
particle numbers shown in Eq. (\ref{number}) as a function of $p_{x}$ with $p_{y}=0$.}
\label{fdlt_h}
\end{figure}

On the other hand, from Eq. ({\ref{berry2D}}) and using integral by parts, $
\mathcal{N}$ can also be given as $\mathcal{N}=1/2\pi \int d^{2}\mathbf{
p\nabla }\cdot \left( \mathbf{p/}\left\vert \mathbf{p}\right\vert
^{2}\right) F_{\mathbf{p}}=\int d^{2}\mathbf{p}\delta \left( \mathbf{p}
\right) F_{\mathbf{p}}=F_{\mathbf{0}}$. This analytical result implies that:
(I) the singular behavior of Berry curvature comes solely from SOC in terms
of the phase factor $\varphi _{\mathbf{p}}$ of the vortex-like solution of
the BdG Hamiltonian; (II) It is the sudden change of $v_{\mathbf{0,}+}^{2}$
instead of SOC that explicitly determines the topological phase transition.
Consequently, there is a sudden change of the ground-state wave-function
associated with the component of triplet pairing of the quasi-particles
denoted by $\beta _{\mathbf{p},+}$ at zero momentum. This is also reflected in the
momentum distribution investigated in \cite{jing} as can be seen from Eq. (\ref{number}) that $
\mathcal{E}_{\mathbf{0},+}/E_{\mathbf{0},+}=sign(h_{c}-h)$ and conclusively proves that
the topological phase transition is directly related to the momentum distribution which can be readily measured in cold atom
experiments. Finally, as expected, this sudden change of the ground-state
wave-function also leads to divergent behavior of FS at $h=h_{c}$. Numerical
result of FS is presented in Fig. \ref{fdlt_h} together with the momentum distribution of the total particle
numbers Eq. (\ref{number}) on both sides of the critical point where the parameters are chosen as $E_{b}/E_{F}=0.5$
and $m\lambda /k_{F}=1$ such that no first order phase transition happens.

In order to obtain the critical behavior of FS around the transition point,
we use another equivalent form of FS \cite{you}: $\chi \left( h\right)
=\sum_{m}\left\vert \left\langle m\right\vert H_{I}\left\vert G\right\rangle
\right\vert ^{2}/\left( E_{m}-E_{g}\right) ^{2}$ where $\left\vert
m\right\rangle $ denotes excited state and $E_{m}$ the excitation energy.
Here $H_{I}=\partial_{h}H$ can be considered as the driving term and $\chi
\left( h \right) $ is directly related with the dynamical response of the
system \cite{shijian}. After direct but lengthy calculations, we obtain FS
as functions of $h$:
\begin{equation}
\chi \left( h\right) =\sum_{\mathbf{p}>0}\left[ \frac{M_{\mathbf{p},+}^{2}}{
E_{ \mathbf{p},+}^{2}}+2\frac{M_{\mathbf{p},0}^{2}}{\left(E_{\mathbf{p}
,+}+E_{\mathbf{p},-}\right) ^{2}}+\frac{M_{\mathbf{p},-}^{2}}{E_{ \mathbf{p}
,-}^{2}}\right]  \label{fsim}
\end{equation}
where matrix elements $M_{\mathbf{p},s}$ and $M_{\mathbf{p},0}$ are given in \cite{supplementray}.

From Eq. ({\ref{fsim}}), it is clear that the divergent behavior solely
comes from the first term where $E_{\mathbf{p},+}=0$ at the transition point
around the Fermi point $p_{x}=p_{y}=0$. Besides, since the gap and chemical
potential vary continuously across the phase transition, derivatives of gap
and chemical potential with respect to $h$ are irrelevant constants for the
critical behavior of FS. Therefore, we can consider $\Delta $, $\mu $ and $h$
as independent controlling parameters and $H_{I}=\partial _{h}H=-\sigma _{z}$
which indicates that FS is directly related to the spin-spin correlation function
and significantly simplifies the matrix elements $M_{s}$ and $M_{0}$. Close
to the critical Zeeman coupling $h_{c}$, the asymptotic form of $M_{\mathbf{
p },+}$ and the excitation spectrum $E_{\mathbf{p},+}$ around the Fermi
points can be given as $M_{\mathbf{p},+}\simeq -\left( \Delta \lambda
/2h_{c}\right) p_{\bot }/E_{\mathbf{p},+}$ and $E_{\mathbf{p},+}^{2}=\Delta ^{2}\lambda ^{2}p_{\perp
}^{2}/h_{c}^{2}+\left\vert h-h_{c}\right\vert ^{2}$
, respectively  \cite{supplementray}. Substituting these asymptotic results into Eq. ({\ref{fsim}}
), we obtain the critical behavior of FS as  \cite{supplementray}
\begin{equation}
\chi \left( h\right) \propto -\ln \left\vert h-h_{c}\right\vert.
\end{equation}

As a byproduct, we conclude that existence of gapless excitation of the
system does not necessarily means divergence of FS and phase transition. For
example, in the absence of SOC, the Hamiltonian still supports gapless
excitation spectrum when $h=h_{c} $ \cite{hui, mann}. However, it does not lead to
a divergent behavior of FS which can be understood in the following aspects.
First, $\partial _{h}H_{0}$ in $H_{I}$ commutes with total Hamiltonian
without SOC, therefore it does not contribute to FS. Second, it is easy to
show that $\partial _{h}H_{BCS}$ only supports pairing excitations with
opposite spins, therefore $M_{\mathbf{p},\pm }^{2}=0$ and only the second
term in Eq. ({\ \ref{fsim}}) is not zero. Finally, without SOC, the
excitation spectrum takes the following form: $E_{\mathbf{p},s}=\sqrt{\left(
\mathbf{p}^{2}/2m-\mu \right) ^{2}+\Delta ^{2}}-sh$ and the combination $E_{
\mathbf{p},+}+E_{\mathbf{p},-}=2 \sqrt{\left( \mathbf{p}^{2}/2m-\mu \right)
^{2}+\Delta ^{2}}$ is always gapped. Therefore,without SOC, the gapless
nature of the excitation spectrum does not manifest itself by causing
divergence of FS and therefore no continuous phase transition.

\textit{Conclusion.}---We investigate the ground-state properties of a
pairing system in the presence of both SOC and Zeeman coupling that supports
non-trivial topological order. In particular, we obtain the analytical result for
the topological invariant which directly relates the topological phase transition
with a sudden change of the BCS-type ground state wave-function at
zero momentum. Furthermore, it conclusively demonstrates that the topological phase transition can be determined by measuring the momentum distribution in cold atomic experiments.
Generalization of this method of evaluating the topological invariant to higher
dimensions and lattice situations will be of great interest. Last but not least, in the absence of Zeeman field without phase transitions, FS shows some new features that are not revealed by other thermodynamic quantities in the BCS-BEC crossover induced by SOC.

\textit{Acknowledgements.}---We acknowledge helpful discussions with Prof.
Gordon Baym, Anthony J. Leggett and S. J. Gu. This work has been supported by the National
Natural Science Foundation of China under Grant 51331006, 11204321, the
National Basic Research Program No. 2010CB934603, the Ministry of Science and
Technology of China and in part by NSF Grants PHY09-69790 and PHY13-05891. K. Z. Zhou also acknowledges China Scholarship Council
for supporting his visit to UIUC where part of this work is carried out.

\begin{widetext}
\section*{Supplementary material for \\
``Fidelity susceptibility and topological phase transition of a two dimensional spin-orbit coupled Fermi superfluid"}

In this supplementary material, we give detailed derivation of Eq. (12) with
explicit form of matrix elements $M_{\mathbf{p},+}$, $M_{\mathbf{p},-}$ and $%
M_{\mathbf{p},0}$ being presented and the critical behavior of fidelity
susceptibility shown in Eq. (13).

\subsection{Derivation of Eq. (12)}

Fidelity susceptibility are defined as

\begin{equation}
\chi =\sum_{m}\frac{\left\vert \left\langle m\right\vert H_{I}\left\vert
G\right\rangle \right\vert ^{2}}{\left( E_{m}-E_{g}\right) ^{2}}  \tag{S1}
\label{fs1}
\end{equation}

with $\left\vert G\right\rangle $ and $\left\vert m\right\rangle $ denote
the ground and excited state, respectively. Here, $H_{I}=\partial H/\partial
h=\sum_{\mathbf{p}>0}\Phi _{\mathbf{p}}^{\dagger }\partial _{h}H_{BdG}\left(
\mathbf{p}\right) \Phi _{\mathbf{p}}+C$ with $\partial _{h}H_{BdG}\left(
\mathbf{p}\right) $ being given as
\begin{equation}
\partial _{h}H_{BdG}\left( \mathbf{p}\right) =\left[
\begin{array}{cccc}
-\partial _{h}\mu -1 & 0 & 0 & -\partial _{h}\Delta \\
0 & -\partial _{h}\mu +1 & \partial _{h}\Delta & 0 \\
0 & \partial _{h}\Delta & \partial _{h}\mu +1 & 0 \\
-\partial _{h}\Delta & 0 & 0 & \partial _{h}\mu -1%
\end{array}%
\right] \text{.}  \tag{S2}  \label{hamI}
\end{equation}

and $C$ being a constant. Since only excited states account for the
summation in Eq. ({\ref{fs1}}), $C$ does not contribute to $\chi $ and can
be neglected in the following discussion.

Using the Bogoliubov transformation described in the main text, $H_{I}$ can
be re-written in terms of the quasi-particle operator $\alpha _{\mathbf{p}%
,s}=u_{\mathbf{p},s}\beta _{\mathbf{p},s}-e^{is\varphi _{\mathbf{p}}}v_{%
\mathbf{p},s}\beta _{-\mathbf{p},s}$.

\begin{equation}
H_{I}=\sum_{\mathbf{p}>0}\left[
\begin{array}{cccc}
\alpha _{-\mathbf{p},+} & \alpha _{\mathbf{p},+}^{\dagger } & \alpha _{-%
\mathbf{p},-} & \alpha _{\mathbf{p},-}^{\dagger }%
\end{array}%
\right] M^{\prime }\left[
\begin{array}{c}
\alpha _{-\mathbf{p},+}^{\dagger } \\
\alpha _{\mathbf{p},+} \\
\alpha _{-\mathbf{p},-}^{\dagger } \\
\alpha _{\mathbf{p},-}%
\end{array}%
\right]  \tag{S3}  \label{hamq}
\end{equation}

with $M^{\prime }$ being the matrix elements. Substituting of Eq. ({\ref%
{hamq}}) into Eq. ({\ref{fs1}}) and using the fact that $\alpha _{\mathbf{p}%
,s}\left\vert G\right\rangle =0$, one can obtain Eq. (12)

\begin{equation}
\chi \left( h\right) =\sum_{\mathbf{p}>0}\left[ \frac{M_{\mathbf{p},+}^{2}}{%
E_{\mathbf{p},+}^{2}}+2\frac{M_{\mathbf{p},0}^{2}}{\left( E_{\mathbf{p}%
,+}+E_{\mathbf{p},-}\right) ^{2}}+\frac{M_{\mathbf{p},-}^{2}}{E_{\mathbf{p}%
,-}^{2}}\right]  \tag{S4}  \label{fs2}
\end{equation}

with matrix elements $M_{\mathbf{p},+}$, $M_{\mathbf{p},-}$ and $M_{\mathbf{p%
},0}$ are given as

\begin{align}
M_{\mathbf{p},+}& =\partial _{h}\mu \left( v_{\mathbf{p}}^{2}-u_{\mathbf{p}%
}^{2}\right) u_{\mathbf{p,}+}v_{\mathbf{p,}+}+\left( 1-2u_{\mathbf{p}}v_{%
\mathbf{p}}\partial _{h}\Delta \right) u_{\mathbf{p,}+}v_{\mathbf{p,}+}\cos
\theta _{\mathbf{p}}+\left( u_{\mathbf{p}}v_{\mathbf{p}}-\frac{\partial
_{h}\Delta }{2}\right) \sin \theta _{\mathbf{p,}}\left( u_{\mathbf{p,}%
+}^{2}-v_{\mathbf{p,}+}^{2}\right) \text{,}  \tag{S5}  \label{m1} \\
M_{\mathbf{p},-}& =\left( u_{\mathbf{p}}^{2}-v_{\mathbf{p}}^{2}\right)
\left( u_{\mathbf{p,}+}v_{\mathbf{p,}-}-u_{\mathbf{p,}-}v_{\mathbf{p,}%
+}\right) \sin \theta _{\mathbf{p}}+\left[ 2u_{\mathbf{p}}v_{\mathbf{p}%
}\partial _{h}\mu -\partial _{h}\Delta \left( u_{\mathbf{p}}^{2}-v_{\mathbf{p%
}}^{2}\right) \cos \theta _{\mathbf{p}}\right] \left( u_{\mathbf{p,}+}u_{%
\mathbf{p,}-}+v_{\mathbf{p,}+}v_{\mathbf{p,}-}\right) \text{,}  \tag{S6}
\label{m2} \\
M_{\mathbf{p},0}& =\partial _{h}\mu \left( v_{\mathbf{p}}^{2}-u_{\mathbf{p}%
}^{2}\right) u_{\mathbf{p,}-}v_{\mathbf{p,}-}-\left( 1+2u_{\mathbf{p}}v_{%
\mathbf{p}}\partial _{h}\Delta \right) u_{\mathbf{p,}-}v_{\mathbf{p,}-}\cos
\theta _{\mathbf{p}}-\left( u_{\mathbf{p}}v_{\mathbf{p}}+\frac{\partial
_{h}\Delta }{2}\right) \sin \theta _{\mathbf{p}}\left( u_{\mathbf{p,}%
-}^{2}-v_{\mathbf{p,}-}^{2}\right) \text{.}  \tag{S7}  \label{m3}
\end{align}

\subsection{Derivation of Eq. (13)}

At the critical point, $h=h_{c}$, the excitation spectrum $E_{\mathbf{p}%
,+}=0 $ at $\mathbf{p=0}$ while $E_{\mathbf{p},-}$ is always gapped. From
Eq. ({\ref{fs2}}), it is clear that the integration is well behaved in the
ultraviolet limit. In the infrared limit, only the first term may cause
divergent behavior of $\chi \left( h_{c}\right) $ because of the gapless
excitation spectrum in the denominator in the $\mathbf{p\rightarrow 0}$
regime.

In order to obtain the asymptotic form of the excitation spectrum, we
investigate the following identity:

\begin{equation}
E_{\mathbf{p},+}^{2}E_{\mathbf{p},-}^{2}=\left( \varepsilon _{\mathbf{p}%
}^{2}+\Delta ^{2}+h^{2}+\left\vert \Gamma _{\mathbf{p}}\right\vert
^{2}\right) ^{2}-4\left( \varepsilon _{\mathbf{p}}^{2}h^{2}+\varepsilon _{%
\mathbf{p}}^{2}\left\vert \Gamma \right\vert ^{2}+\Delta ^{2}h^{2}\right)
\text{.}  \tag{S8}
\end{equation}

Around the critical point $h=h_{c}+\eta $ with $\eta $ being small, Taylor
expansion of the above identity to the lowest (second) order in $p_{\perp }$
and $\eta $, we obtain

\begin{equation}
E_{\mathbf{p},+}^{2}E_{\mathbf{p},-}^{2}\rightarrow 4\Delta ^{2}\lambda
^{2}p_{\perp }^{2}+4h_{c}^{2}\eta ^{2}  \tag{S9}  \label{epen}
\end{equation}

Since $E_{\mathbf{p},-}$ is always gapped, in the zeroth order,

\begin{equation}
E_{\mathbf{p},-}^{2}\rightarrow \left( \sqrt{\mu ^{2}+\Delta ^{2}}+h\right)
^{2}=4h_{c}^{2}  \tag{S10}  \label{en}
\end{equation}

Together with Eq. ({\ref{epen}}) and Eq. ({\ref{en}}), we obtain the
asymptotic form of the excitation spectrum $E_{\mathbf{p},+}^{2}$

\begin{equation}
E_{\mathbf{p},+}^{2}=\frac{\Delta ^{2}\lambda ^{2}}{h_{c}^{2}}p_{\perp
}^{2}+\eta ^{2}  \tag{S11}  \label{ep}
\end{equation}

As have been stated in the main text, for the topological phase transition
considered in this Letter, gap parameter $\Delta $ and chemical potential $%
\mu $ as functions of Zeeman field $h$ are continuous and their derivatives
(appearing in the enumerator) are irrelevant to the divergent behavior of $%
\chi \left( h\right) $. Therefore, matrix element $M_{\mathbf{p},+}$ can be
reduced to%
\begin{align}
M_{\mathbf{p},+} =&u_{\mathbf{p,}+}v_{\mathbf{p,}+}\cos \theta _{\mathbf{p}%
}+u_{\mathbf{p}}v_{\mathbf{p}}\sin \theta _{\mathbf{p,}}\left( u_{\mathbf{p,}%
+}^{2}-v_{\mathbf{p,}+}^{2}\right)  \notag \\
&=\frac{1}{2}\frac{\Delta \sin \theta _{\mathbf{p}}\cos \theta _{\mathbf{p}}%
}{E_{\mathbf{p,}+}}+\frac{1}{2}\frac{\Delta \left\vert \cos \theta _{\mathbf{%
p}}\right\vert \sin \theta _{\mathbf{p}}}{E_{\mathbf{p}}}\frac{\mathcal{E}_{%
\mathbf{p,}+}}{E_{\mathbf{p,}+}}  \notag \\
&=\frac{1}{2}\frac{\Delta \sin \theta _{\mathbf{p}}\left\vert \cos \theta _{%
\mathbf{p}}\right\vert }{E_{\mathbf{p,}+}}\left( \frac{\mathcal{E}_{\mathbf{%
p,}+}}{E_{\mathbf{p}}}-1\right) \text{.}  \tag{S12}  \label{emap}
\end{align}

Around the critical point $h_{c}$, $\sin \theta _{\mathbf{p}}\propto
p_{\perp }$, $\left\vert \cos \theta _{\mathbf{p}}\right\vert =1$ and $%
\mathcal{E}_{\mathbf{p,}+}\propto p_{\perp }$. Therefore, at first order in $%
p_{\perp }$, $M_{\mathbf{p},+}$ can be approximated as

\begin{equation}
M_{\mathbf{p},+}\rightarrow -\frac{1}{2}\frac{\Delta \sin \theta }{E_{%
\mathbf{p,}+}}=-\frac{1}{2}\frac{\Delta \lambda p_{\perp }}{E_{\mathbf{p,}%
+}h_{c}}\text{.}  \tag{S13}
\end{equation}

Together with Eq. ({\ref{ep}}), we obtain the asymptotic form of $M_{\mathbf{%
p},+}$ as

\begin{equation}
M_{\mathbf{p},+}\rightarrow -\frac{1}{2}\frac{\Delta \lambda p_{\perp }}{%
\sqrt{\frac{\Delta ^{2}\lambda ^{2}}{h_{c}^{2}}p_{\perp }^{2}+\left\vert
h-h_{c}\right\vert ^{2}}h_{c}}\text{.}  \tag{S14}  \label{map}
\end{equation}

Substituting Eq. ({\ref{map}}) and Eq. ({\ref{ep}}) into the first term in
Eq. ({\ref{fs2}}), we obtain

\begin{equation}
\chi \left( h\right) =\frac{1}{4}\sum_{\mathbf{p}>0}\frac{\Delta ^{2}\lambda
^{2}p_{\perp }^{2}}{h_{c}\left( \frac{\Delta ^{2}\lambda ^{2}}{h_{c}^{2}}%
p_{\perp }^{2}+\left\vert h-h_{c}\right\vert ^{2}\right) ^{2}}  \tag{S15}
\end{equation}

As usual, summation over momentum is replaced by integration in the
continuum limit and we only consider the infrared regime of the momentum
integration which causes divergence of $\chi \left( h\right) $. Based on
these assumptions, we obtain

\begin{align}
\chi \left( h\right) \propto &\int_{0}p_{\perp }dp_{\perp }\frac{\Delta
^{2}\lambda ^{2}p_{\perp }^{2}}{h_{c}^{2}\left( \frac{\Delta ^{2}\lambda ^{2}%
}{h_{c}^{2}}p_{\perp }^{2}+\left\vert h-h_{c}\right\vert ^{2}\right) ^{2}}%
\propto \int_{0}dx\frac{\frac{\Delta ^{2}\lambda ^{2}}{h_{c}^{2}}x}{\left(
\frac{\Delta ^{2}\lambda ^{2}}{h_{c}^{2}}x+\left\vert h-h_{c}\right\vert
^{2}\right) ^{2}}  \notag \\
&\propto \int dx\frac{x}{\left( x+\left\vert h-h_{c}\right\vert ^{2}\right)
^{2}}\simeq -1-\ln \left\vert h-h_{c}\right\vert ^{2}  \notag \\
&\propto -\ln \left\vert h-h_{c}\right\vert  \tag{S16}  \label{fs3}
\end{align}
where we neglect the exact value of the coefficient which in general depends
on the derivative of $\Delta $ and $\mu $ with respect to $h$.

Finally, we conclude that the terms depending on the derivative of $\Delta $
and $\mu $ with respect to $h$ do not affect the logarithmic divergence of $%
\chi $. First, for $p_{\perp }\rightarrow 0$, $v_{\mathbf{p}}^{2}-u_{\mathbf{%
p}}^{2}$, $1-2u_{\mathbf{p}}v_{\mathbf{p}}\partial _{h}\Delta $, $\cos
\theta _{\mathbf{p}}$ and $u_{\mathbf{p}}v_{\mathbf{p}}-\frac{\partial
_{h}\Delta }{2}$ are constants and $u_{\mathbf{p,}+}v_{\mathbf{p,}+}\propto
p_{\perp }/E_{\mathbf{p,}+}$ and $\sin \theta _{\mathbf{p}}\propto p_{\perp }
$. Second, the third term is second order in $p_{\perp }$ as can be seen
from Eq. ({\ref{emap}}). Therefore, in the lowest order in $p_{\perp }$, $M_{%
\mathbf{p},+}\propto p_{\perp }/E_{\mathbf{p,}+}$ which leads to the
logarithmic divergence of $\chi \left( h\right) $ as can be seen from Eq. ({%
\ref{fs3}}).
\end{widetext}

\end{document}